\documentclass[Physsubmission, Phys]{SciPost}

\hypersetup{
    colorlinks,
    linkcolor={red!50!black},
    citecolor={blue!50!black},
    urlcolor={blue!80!black}
}

\usepackage[bitstream-charter]{mathdesign}
\usepackage{placeins}
\DeclareSymbolFont{usualmathcal}{OMS}{cmsy}{m}{n}
\DeclareSymbolFontAlphabet{\mathcal}{usualmathcal}
 
\usepackage[absolute,overlay]{textpos}

\begin{document}
\begin{textblock*}{5cm}(15cm,0.7cm) 
\textbf{ MS-TP-21-18}
\end{textblock*}
\begin{center}{\Large \textbf{
Constraining the nuclear gluon PDF with inclusive hadron production data\\
}}\end{center}

\begin{center}
P.~Duwentäster\textsuperscript{1$\star$},
L.~A.~Husová\textsuperscript{2},
T.~Ježo\textsuperscript{3},
M.~Klasen\textsuperscript{1},
K.~Kovařík\textsuperscript{1},
A.~Kusina\textsuperscript{4},
K.~F.~Muzakka\textsuperscript{1},
F.~I.~Olness\textsuperscript{5},
I.~Schienbein\textsuperscript{6} and
J.Y.~Yu\textsuperscript{5}
\end{center}

\begin{center}
{\bf 1} Institut für Theoretische Physik, Westfälische Wilhelms-Universität Münster,Wilhelm-Klemm-Straße 9, D-48149 Münster, Germany
\\
{\bf 2} Institut für Kernphysik, Westfälische Wilhelms-Universität Münster,Wilhelm-Klemm-Straße 9, D-48149 Münster, Germany
\\
{\bf 3} Institute  for  Theoretical  Physics,  KIT,  D-76131  Karlsruhe,  Germany
\\
{\bf 4} Institute  of  Nuclear  Physics  Polish  Academy  of  Sciences,  PL-31342  Krakow,  Poland
\\
{\bf 5} Southern  Methodist  University,  Dallas,  TX  75275,  USA
\\
{\bf 6} Laboratoire de Physique Subatomique et de Cosmologie, Universit´e Grenoble-Alpes,
CNRS/IN2P3, 53 avenue des Martyrs, 38026 Grenoble, France
\\
* pit.duw@uni-muenster.de
\end{center}

\begin{center}
\today
\end{center}


\definecolor{palegray}{gray}{0.95}
\begin{center}
\colorbox{palegray}{
  \begin{tabular}{rr}
  \begin{minipage}{0.1\textwidth}
    \includegraphics[width=22mm]{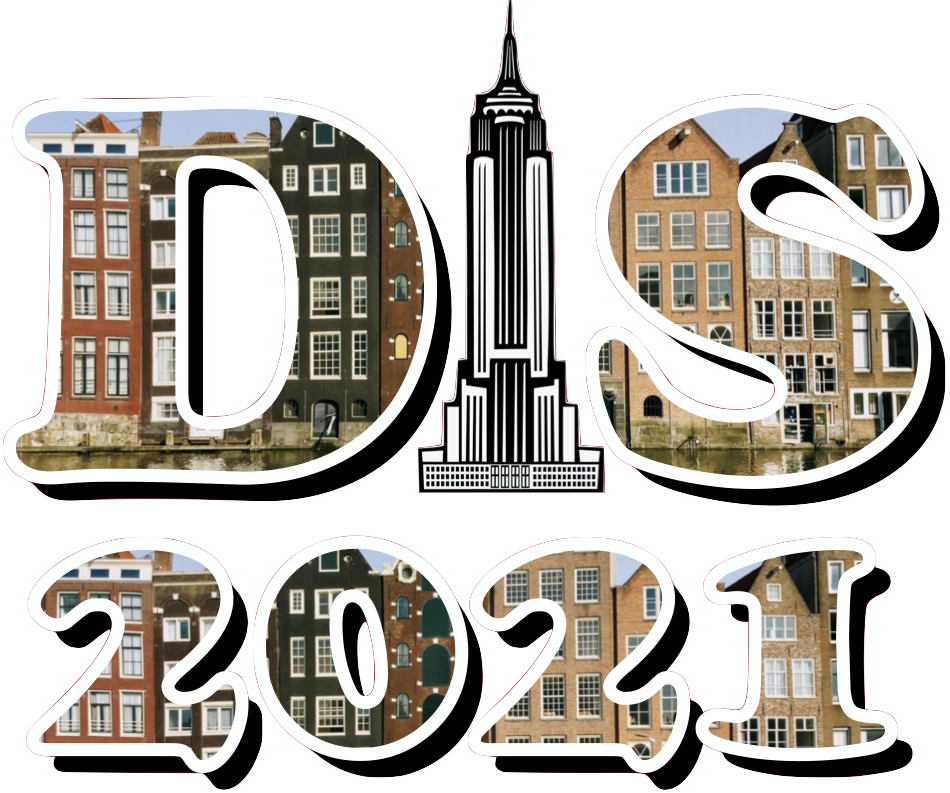}
  \end{minipage}
  &
  \begin{minipage}{0.75\textwidth}
    \begin{center}
    {\it Proceedings for the XXVIII International Workshop\\ on Deep-Inelastic Scattering and
Related Subjects,}\\
    {\it Stony Brook University, New York, USA, 12-16 April 2021} \\
    \doi{10.21468/SciPostPhysProc.?}\\
    \end{center}
  \end{minipage}
\end{tabular}
}
\end{center}

\section*{Abstract}
{\bf
The nuclear parton distribution functions (nPDFs) of gluons are known
to be difficult to determine with fits of deep inelastic scattering (DIS)
and Drell-Yan (DY) data alone. Therefore, the nCTEQ15 analysis of nuclear PDFs
added inclusive neutral pion production data from RHIC to help in constraining the gluon. 
In this analysis, we present a new global analysis of nuclear PDFs based
on a much larger set of single inclusive light hadron data from RHIC
and the LHC. Using our new nCTEQ code (\texttt{nCTEQ++}) with an optimized version of INCNLO we study systematically the limitations of the theory and the 
impact of the fragmentation function uncertainty.
}


\section{Introduction}
\label{sec:intro}
The QCD parton model has been used with great success to make predictions for experiments at SLAC, HERA, TeVatron, RHIC and LHC. However, the parton distribution functions necessary for this cannot be determined perturbatively and the current best option is to fit them in global QCD analyses. Especially in the case of nuclear PDFs, analyses including only deep inelastic scattering  (DIS)  and  the  Drell-Yan (DY) process data struggle to put strong constraints on the gluon, since these processes are not directly sensitive to gluons at leading order. To address this problem we investigate the impact of single inclusive hadron production data, which is dominated by gluon initiated subprocesses in the kinematic range of the available data.


\subsection{nCTEQ framework}
The nCTEQ framework expands the global analyses of proton PDFs towards full nuclei. We parameterize the effective proton PDFs at an initial scale $Q_0=1.3$\,GeV as
\begin{align*}
    xf_i^{p/A}(x,Q_0)=c_0x^{c_1}(1-x)^{c_2}e^{c_3x}(1+e^{c_4}x)^{c_5} 
\quad \quad  \text{with} \quad \quad
    c_k(A) \equiv p_{k}+ a_{k}(1-A^{b_{k}}),
\end{align*}
where the $p_k$ parameters characterize the proton baseline, $a_k$ the nuclear modification and $b_k$ the nuclear $A$ dependence. The full nuclear PDFs are then calculated as the sum of $Z$ effective protons and $A-Z$ effective neutrons assuming isospin symmetry. A full overview of the framework is provided in \cite{Kovarik:2015cma}. We open the same 19 parameters as in nCTEQ15WZ~\cite{Kusina:2020lyz} in our new fits. The cross sections for inclusive hadron production are calculated via a modified version of INCNLO~\cite{INCNLO} that uses precomputed grids for the convolutions without the nuclear PDFs for improved speed. 

\subsection{Proton-proton baseline and scale dependence}
Before we can include any data of nuclear ratios in our fit we need to make sure that the proton-proton baseline can be accurately described by our framework. 

The computed cross sections depend on two choices: The first choice is the Fragmentation Function (FF) describing the fragmentation of the initially produced particle into the final state hadron. These fragmentation functions cannot be calculated perturbatively and need to be fitted similarly to PDFs. The uncertainty introduced by this is largely mitigated by the fact that nuclear ratios are used in the fit. Additionally, we compute the uncertainties using the eigenvectors provided with the FFs and add this to the systematic uncertainties of the data to account for any remaining effects.
In this analysis, we limit ourselves to DSS\footnote{DSS14~\cite{deFlorian:2014xna} for pions and DSS17~\cite{deFlorian:2017lwf} for kaons} FFs~\cite{deFlorian:2014xna, deFlorian:2017lwf} in this analysis and refer to \cite{Duwentaster:2021ioo} for a comparison with a variety of other FFs.

\begin{figure*}[h!]
		\centering
		\includegraphics[width=0.99\textwidth]{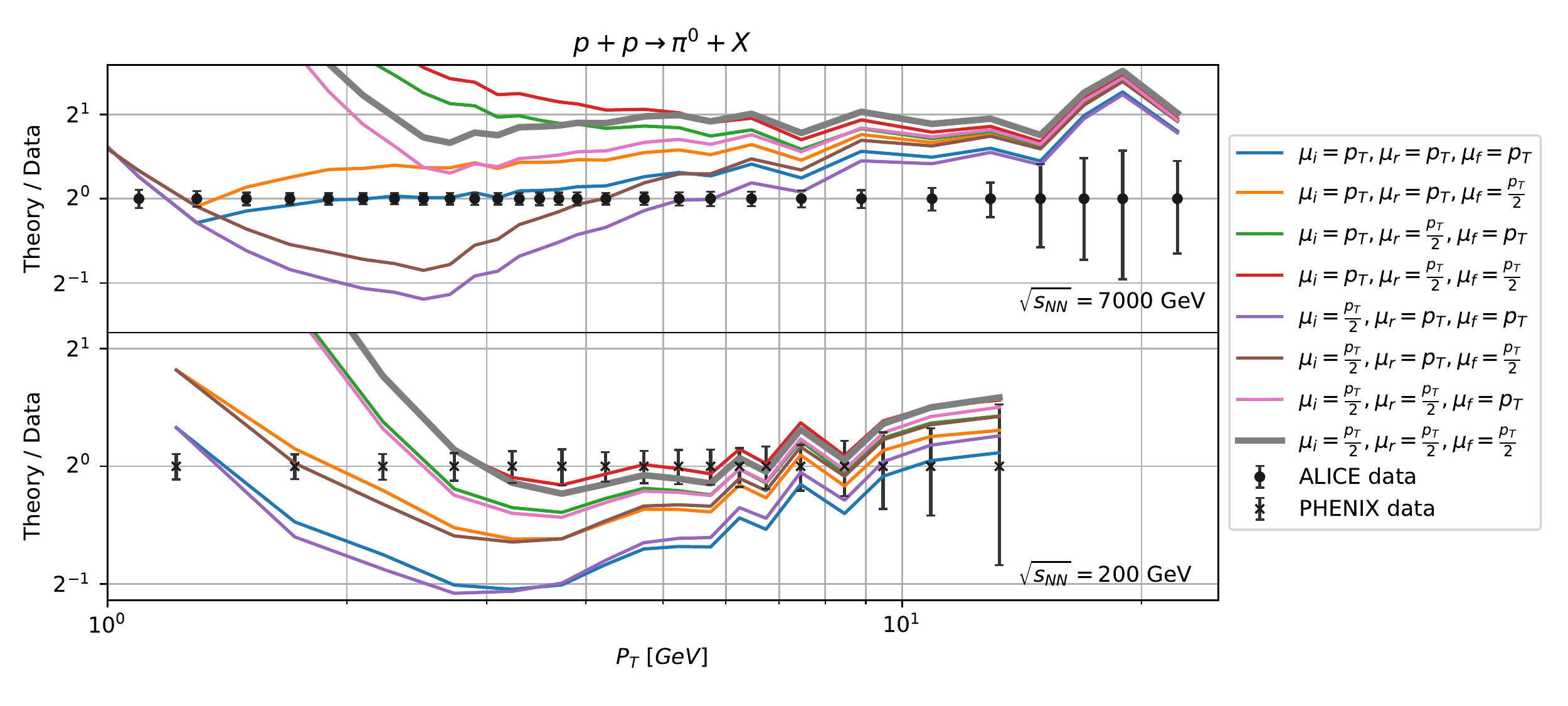}
	\caption{Comparison of different scale choices for predictions of ${p+p\rightarrow\pi^0+X}$ shown as the ratio of theory over data for PHENIX~\cite{Adler:2003pb} in the upper panel
	and \mbox{ALICE~\cite{Abelev:2012cn}} in the lower panel. The theory predictions are calculated using nCTEQ15 proton PDFs with DSS FFs.}
	\label{fig:scale}
\end{figure*}

The second choice are the scales for the initial factorization, renormalization and final factorization. The effect of these can be seen in Fig.~\ref{fig:scale} which shows the ratio of theory prediction over data for $p+p\rightarrow \pi^0+X$ for different choices of the three scales. Setting all scales equal to $p_T/2$ allows us to describe all data  above $p_T=3$\,GeV if a normalization factor is introduced. Since the data in the fits will be ratios the normalization factors will cancel.
    
\subsection{Available data}
Table~\ref{tab:data_table} lists the nuclear data sets included in this analysis. The data from STAR and PHENIX is taken at $\sqrt{s_{NN}}=200$\,GeV and the data from ALICE at $5.02$\,TeV and $8.16$\,TeV. Cutting the data below $p_T=3$\,GeV leaves us with 77 (out of 174) ALICE and 32 (out of 77) RHIC data points. We exclude data of other light hadrons from the analysis due to their larger uncertainties, but the effect of including eta meson production is investigated in \cite{Duwentaster:2021ioo}.

\begin{table}[h!]
    	\caption{Overview of the available data sets and number of data points after / before cuts.}
     \begin{minipage}{.45\linewidth}
        \renewcommand{\arraystretch}{1.2}  
    	\centering	
    	\begin{tabular}{|c|c|c|c|}
    		\hline 
    		Data set & Ref. & Observ. & No. points \\ 
    		\hline
    		\hline
    		PHENIX $\pi^0$ &~\cite{Adler:2006wg}  & $R_{d\mathrm{Au}}$ & 17/21  \\ 
    		\hline 
    		PHENIX $\pi^\pm$ &~\cite{Adare:2013esx}   & $R_{d\mathrm{Au}}$ & 0/20  \\ 
    		\hline 
    		PHENIX $K^\pm$ &~\cite{Adare:2013esx} & $R_{d\mathrm{Au}}$ & 0/15  \\ 
    		\hline 
    		STAR$ \pi^0$ &~\cite{Abelev:2009hx} & $R_{d\mathrm{Au}}$ & 9/13 \\ 
    		\hline 
    		STAR $\pi^\pm$ &~\cite{Adams:2006nd} & $R_{d\mathrm{Au}}$ & 8/23 \\ 
    		\hline 
    	\end{tabular}     	
    	\label{tab:data_table}
     \end{minipage}
     \hfill
     \begin{minipage}{.45\linewidth}
        \renewcommand{\arraystretch}{1.2}  
    	\centering	
    	\vspace{-0.6cm}
    	\begin{tabular}{|c|c|c|c|}
    		\hline 
    		Data set & Ref. & Observ. & No. points \\ 
    		\hline
    		\hline
    		ALICE 5\,TeV $\pi^0$ &~\cite{Acharya:2018hzf} & $R_{p\mathrm{Pb}}$ & 15/31 \\ 
    		\hline 
    		ALICE 5\,TeV $\pi^\pm$ &~\cite{Adam:2016dau} & $R_{p\mathrm{Pb}}$ & 22/58 \\ 
    		\hline 
    		ALICE 5\,TeV $K^\pm$ &~\cite{Adam:2016dau} & $R_{p\mathrm{Pb}}$ & 22/58 \\ 
    		\hline 
    		ALICE 8\,TeV $\pi^0$ &~\cite{Acharya:2021yrj} & $R_{p\mathrm{Pb}}$ & 19/30 \\ 
    		\hline 
    	\end{tabular}     	
    	\label{tab:data_table}
     \end{minipage}
\end{table}
\section{PDF fits with SIH data}
The fits are performed with the same data and cuts as nCTEQ15(WZ) for the DIS, DY and WZ data, while adding the new SIH data with a cut at $p_T = 3$\,GeV. We use DSS fragmentation functions and add their uncertainty to the systematic uncertainty of the data. The normalizations of the SIH data sets are fitted according to the prescription given in \cite{DAgostini:1993arp}.
\subsection{Resulting PDFs}
\begin{figure*}[h!]
	\centering
	\includegraphics[width=0.48\textwidth]{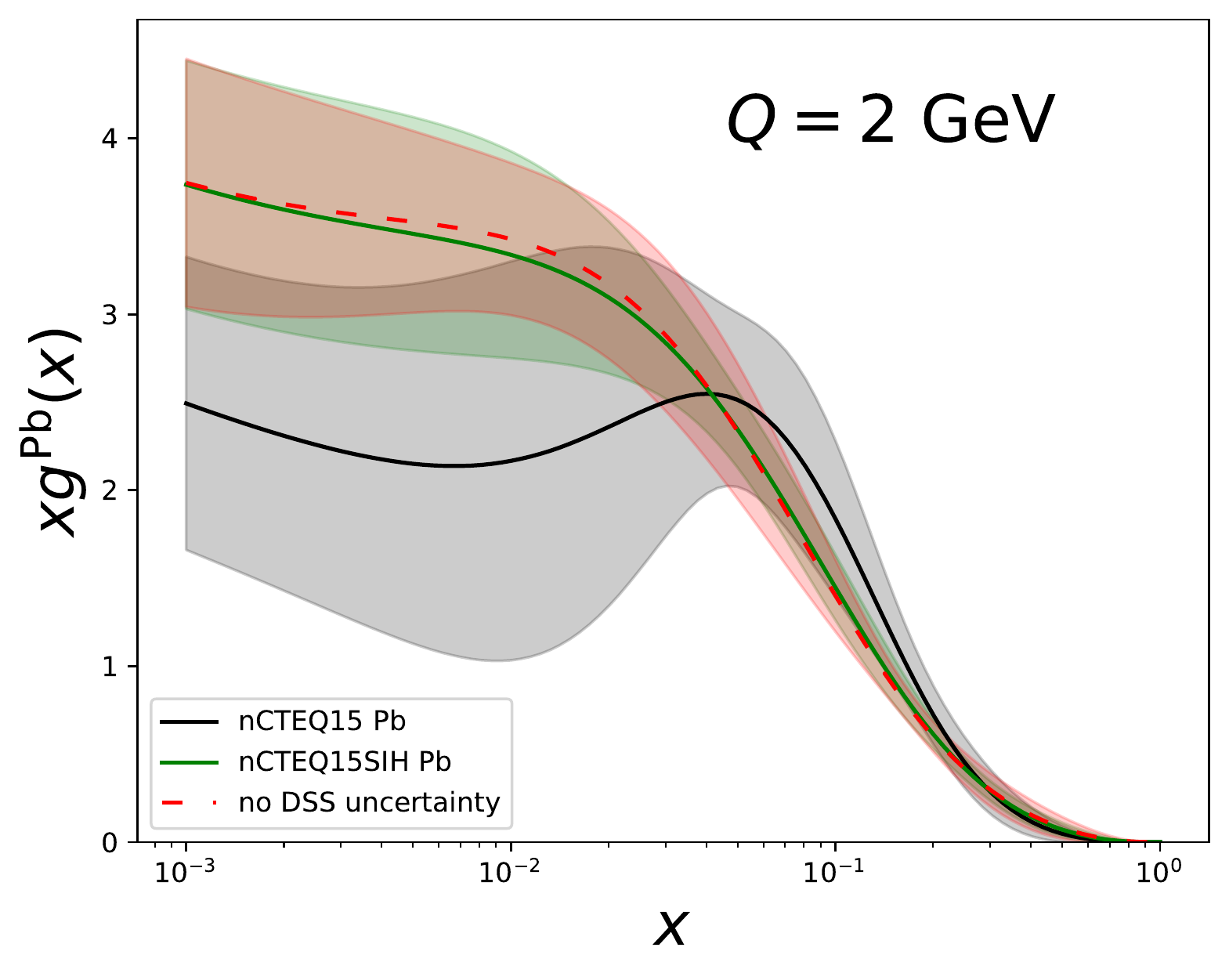}
	\includegraphics[width=0.48\textwidth]{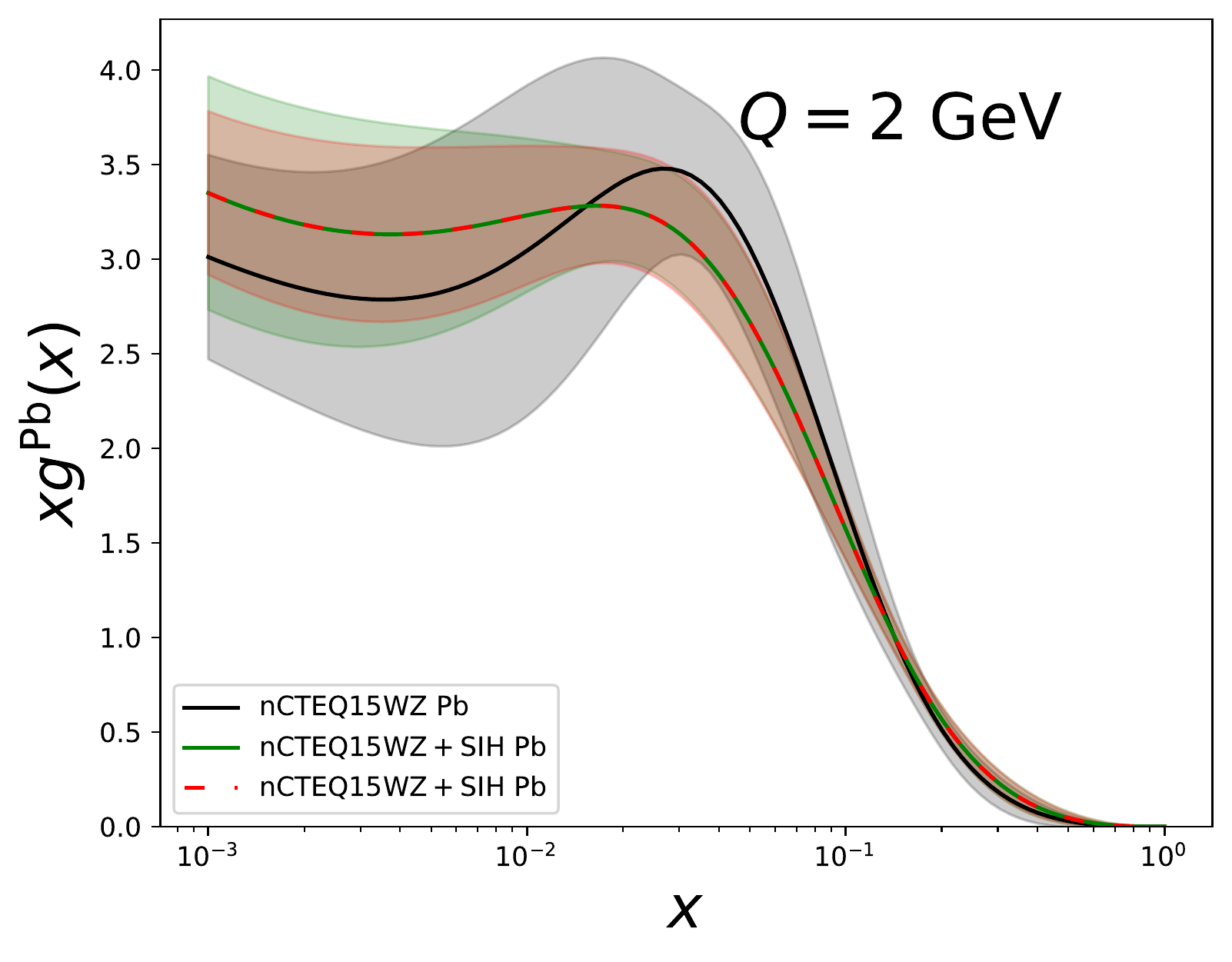}
	\caption{Gluon PDFs in lead. The baseline nCTEQ15 (left) / nCTEQ15WZ (right) is shown in black, the fit with unmodified data is shown in red, and the fit with added uncertainties from the DSS FFs is shown in green.
	}
	\label{fig:main_pdf_WZ}
\end{figure*}

Figure \ref{fig:main_pdf_WZ} shows the gluon PDFs in lead for the fits with SIH data and their respective baselines. 
In both cases the inclusion of SIH data raises the low-$x$ gluon while causing a suppression in the medium-$x$ region. This effect is less pronounced in the nCTEQ15WZ based fit since the WZ production data puts additional constraints on the gluon. Adding the DSS uncertainties to the systematic unceratinties of the data leads to a negligible shift in the fits' central values, but does increase the final uncertainty band slightly.
Other flavors do not change much in terms of central values but the uncertainties in the nCTEQ15 based fits increase due to the newly opened strange quark parameters. A more detailed discussion of those flavors is provided in \cite{Duwentaster:2021ioo}.
\subsection{Quality of the fits}
To investigate how well the fits describes the data we take a look at the resulting $\chi^2/N_{dof}$ values for each process. Table~\ref{tab:chi2table_mainfits} shows these $\chi^2/N_{dof}$ values and the total for the two baseline fits and the fits including the DSS uncertainties. We see that nCTEQ15 does not deliver a good description of the WZ data. The value of 1.23 for SIH data does not look bad at first sight, but including the data sets in the fit shows a tremendous improvement to 0.38 while also improving the description of the WZ production data. Similarly, adding only the WZ data to the nCTEQ15 fit yields a large improvement in the $\chi^2/N_{dof}$ for this process and some improvement for SIH data suggesting a similar impact on the gluon PDF. Finally, including both data sets allows a good description of all processes and the best total $\chi^2/N_{dof}=0.85$.
A more thorough examination of the fit quality in terms of correlations with individual data sets and a comparison with other FFs can be found in \cite{Duwentaster:2021ioo}.
\begin{table*}[h!]	
\renewcommand{\arraystretch}{1.2}  
	\caption{$\chi^2/N_{dof}$ for the individual processes {DIS, DY, WZ, SIH}, and the total. 
	\mbox{Excluded} processes are shown in parentheses. (Note that nCTEQ15 and nCTEQ15WZ include the neutral pion data from RHIC.)
	}
	\centering	
	\begin{tabular}{|c|c|c|c|c|c|}
    		\hline 
    		\multicolumn{6}{|c|}{$\chi^2/N_{d.o.f.}$ for individual processes} \\ 
    		\hline
    		 &  DIS  & DY & WZ & SIH & \textbf{Total}\\ 
    		\hline 
    		\hline
    		nCTEQ15         & 0.86 & 0.78 & (3.74) & (1.23) & \textbf{1.28} \\ 
    		\hline 
    		nCTEQ15+SIH     & 0.87 & 0.72 & (2.32) & 0.38 & \textbf{1.00} \\ 
    		\hline 
    		nCTEQ15WZ       & 0.90 & 0.78 & 0.90 & (0.81) & \textbf{0.90} \\ 
    		\hline 
    		nCTEQ15WZ+SIH   & 0.91 & 0.77 & 1.02 & 0.41 & \textbf{0.85} \\ 
    		\hline 
    	\end{tabular}    
	\label{tab:chi2table_mainfits}
\end{table*}

\section{Conclusion}
New data on single inclusive hadron (SIH) production from ALICE and RHIC has been incorporated into our PDF analysis. 
We investigated the fragmentation functions uncertainties and choice of scales, and identified a $p_T$ region where reliable perturbative predictions can be made to help constrain the PDFs. 
The resulting impact is consistent with that of the colorless weak boson production data introduced in the nCTEQ15WZ analysis. 
Good $\chi^2/N_{dof}$ values are obtained for all data sets and a noticeable impact on the gluon PDF at low to medium $x$ is observed despite the restrictive $p_T$ cut.
The gluon distribution flattens out in the region $x\approx0.05$ and the uncertainties are reduced in this region.

The obtained nCTEQ15WZ+SIH PDFs for a selection of nuclei will be available through LHAPDF and others can be obtained upon request.
\section*{Acknowledgements}


\paragraph{Funding information}
The work of P.D., M.K.\ and K.K.\ was funded by the Deutsche Forschungsgemeinschaft (DFG, German Research Foundation) – project-id 273811115 – SFB 1225. 
L.A.H., M.K., K.K.\ and K.F.M.\ also acknowledge support of the DFG through the Research Training Group GRK 2149. 
A.K. acknowledges the support of Narodowe Centrum Nauki under Grant No. 2019/34/E/ST2/00186.
The work of T.J.\ was supported by the Deutsche Forschungsgemeinschaft (DFG, German Research Foundation) under grant 396021762 - TRR 257. 
F.O.\ acknowledges support through US DOE grant  No.~DE-SC0010129,
and the National Science Foundation under Grant No.~NSF PHY-1748958.            
The work of I.S.\ was supported by the French CNRS via the IN2P3 project GLUE@NLO.





\FloatBarrier
\bibliography{refs.bib,extra.bib}

\begin{thebibliography}{10}
\providecommand{\url}[1]{\texttt{#1}}
\providecommand{\urlprefix}{URL }
\expandafter\ifx\csname urlstyle\endcsname\relax
  \providecommand{\doi}[1]{doi:\discretionary{}{}{}#1}\else
  \providecommand{\doi}{doi:\discretionary{}{}{}\begingroup
  \urlstyle{rm}\Url}\fi
\providecommand{\eprint}[2][]{\url{#2}}

\bibitem{Kovarik:2015cma}
K.~Kovarik \emph{et~al.},
\newblock \emph{{nCTEQ15 - Global analysis of nuclear parton distributions with
  uncertainties in the CTEQ framework}},
\newblock Phys. Rev. D \textbf{93}(8), 085037 (2016),
\newblock \doi{10.1103/PhysRevD.93.085037},
\newblock \eprint{1509.00792}.

\bibitem{Kusina:2020lyz}
A.~Kusina \emph{et~al.},
\newblock \emph{{Impact of LHC vector boson production in heavy ion collisions
  on strange PDFs}},
\newblock Eur. Phys. J. C \textbf{80}(10), 968 (2020),
\newblock \doi{10.1140/epjc/s10052-020-08532-4},
\newblock \eprint{2007.09100}.

\bibitem{INCNLO}
M.~Werlen,
\newblock \emph{{INCNLO}-direct photon and inclusive hadron production code
  website},
\newblock \urlprefix\url{{http://lapth.cnrs.fr/PHOX_FAMILY}},
\newblock {Version 1.4. http://lapth.cnrs.fr/PHOX\_FAMILY}.

\bibitem{deFlorian:2014xna}
D.~de~Florian, R.~Sassot, M.~Epele, R.~J. Hern\'andez-Pinto and M.~Stratmann,
\newblock \emph{{Parton-to-Pion Fragmentation Reloaded}},
\newblock Phys. Rev. D \textbf{91}(1), 014035 (2015),
\newblock \doi{10.1103/PhysRevD.91.014035},
\newblock \eprint{1410.6027}.

\bibitem{deFlorian:2017lwf}
D.~de~Florian, M.~Epele, R.~J. Hernandez-Pinto, R.~Sassot and M.~Stratmann,
\newblock \emph{{Parton-to-Kaon Fragmentation Revisited}},
\newblock Phys. Rev. D \textbf{95}(9), 094019 (2017),
\newblock \doi{10.1103/PhysRevD.95.094019},
\newblock \eprint{1702.06353}.

\bibitem{Duwentaster:2021ioo}
P.~Duwent\"aster, L.~A. Husov\'a, T.~Je\v{z}o, M.~Klasen, K.~Kova\v{r}\'\i{}k,
  A.~Kusina, K.~F. Muzakka, F.~I. Olness, I.~Schienbein and J.~Y. Yu,
\newblock \emph{{Impact of inclusive hadron production data on nuclear gluon
  PDFs}}  (2021),
\newblock \eprint{2105.09873}.

\bibitem{Adler:2003pb}
S.~S. Adler \emph{et~al.},
\newblock \emph{{Mid-rapidity neutral pion production in proton proton
  collisions at $\sqrt{s}$ = 200-GeV}},
\newblock Phys. Rev. Lett. \textbf{91}, 241803 (2003),
\newblock \doi{10.1103/PhysRevLett.91.241803},
\newblock \eprint{hep-ex/0304038}.

\bibitem{Abelev:2012cn}
B.~Abelev \emph{et~al.},
\newblock \emph{{Neutral pion and $\eta$ meson production in proton-proton
  collisions at $\sqrt{s}=0.9$ TeV and $\sqrt{s}=7$ TeV}},
\newblock Phys. Lett. B \textbf{717}, 162 (2012),
\newblock \doi{10.1016/j.physletb.2012.09.015},
\newblock \eprint{1205.5724}.

\bibitem{Adler:2006wg}
S.~S. Adler \emph{et~al.},
\newblock \emph{{Centrality dependence of $\pi^0$ and $\eta$ production at
  large transverse momentum in $\sqrt{s}_{NN}$ = 200 GeV d{+}Au collisions}},
\newblock Phys. Rev. Lett. \textbf{98}, 172302 (2007),
\newblock \doi{10.1103/PhysRevLett.98.172302},
\newblock \eprint{nucl-ex/0610036}.

\bibitem{Adare:2013esx}
A.~Adare \emph{et~al.},
\newblock \emph{{Spectra and ratios of identified particles in Au+Au and $d$+Au
  collisions at $\sqrt{s_{NN}}=200$ GeV}},
\newblock Phys. Rev. C \textbf{88}(2), 024906 (2013),
\newblock \doi{10.1103/PhysRevC.88.024906},
\newblock \eprint{1304.3410}.

\bibitem{Abelev:2009hx}
B.~I. Abelev \emph{et~al.},
\newblock \emph{{Inclusive $\pi^0$, $\eta$, and direct photon production at
  high transverse momentum in $p+p$ and $d+$Au collisions at
  $\sqrt{s_{NN}}=200$ GeV}},
\newblock Phys. Rev. C \textbf{81}, 064904 (2010),
\newblock \doi{10.1103/PhysRevC.81.064904},
\newblock \eprint{0912.3838}.

\bibitem{Adams:2006nd}
J.~Adams \emph{et~al.},
\newblock \emph{{Identified hadron spectra at large transverse momentum in
  p{+}p and d{+}Au collisions at $\sqrt{s}_{NN}$ = 200 GeV}},
\newblock Phys. Lett. B \textbf{637}, 161 (2006),
\newblock \doi{10.1016/j.physletb.2006.04.032},
\newblock \eprint{nucl-ex/0601033}.

\bibitem{Acharya:2018hzf}
S.~Acharya \emph{et~al.},
\newblock \emph{{Neutral pion and $\eta$ meson production in p-Pb collisions at
  $\sqrt{s_\mathrm{NN}} = 5.02$ TeV}},
\newblock Eur. Phys. J. C \textbf{78}(8), 624 (2018),
\newblock \doi{10.1140/epjc/s10052-018-6013-8},
\newblock \eprint{1801.07051}.

\bibitem{Adam:2016dau}
J.~Adam \emph{et~al.},
\newblock \emph{{Multiplicity dependence of charged pion, kaon, and
  (anti)proton production at large transverse momentum in p-Pb collisions at
  $\mathbf{\sqrt{{\textit s}_{\rm NN}}}$ = 5.02 TeV}},
\newblock Phys. Lett. B \textbf{760}, 720 (2016),
\newblock \doi{10.1016/j.physletb.2016.07.050},
\newblock \eprint{1601.03658}.

\bibitem{Acharya:2021yrj}
S.~Acharya \emph{et~al.},
\newblock \emph{{Nuclear modification factor of light neutral-meson spectra up
  to high transverse momentum in p-Pb collisions at $\sqrt{s_{NN}}$ = 8.16
  TeV}}  (2021),
\newblock \eprint{2104.03116}.

\bibitem{DAgostini:1993arp}
G.~D'Agostini,
\newblock \emph{{On the use of the covariance matrix to fit correlated data}},
\newblock Nucl. Instrum. Meth. A \textbf{346}, 306 (1994),
\newblock \doi{10.1016/0168-9002(94)90719-6}.

\end{thebibliography}

\nolinenumbers

\end{document}